\documentclass[prl,reprint, twocolumn,showpacs,preprintnumbers,amsmath,amssymb,nofootinbib,floatfix]{revtex4-1} 
\usepackage{graphicx}

\usepackage{mathrsfs}
\usepackage{hyperref}

\usepackage{slashed}

\usepackage{xcolor}

\usepackage{gensymb}

\usepackage{amsmath}
\usepackage{empheq}
\allowdisplaybreaks[4]

\usepackage{bm}

\usepackage{multirow}
\usepackage{booktabs}
\usepackage{placeins}

\usepackage[normalem]{ulem}     

\def \matrix #1 {\left(\begin{array}{cc} #1 \end{array}\right)}

\def \e  {\mathop{\rm e}\nolimits}

\def\II{\hbox{{1}\kern-.25em\hbox{l}}}

\newcommand \widebar [1] {\overline{#1}}

\definecolor{RED}{rgb}{1,0,0}\definecolor{BLUE}{rgb}{0,0,1} 

\begin{document}

\title{Three-Loop QCD  Evolution for Nonforward  Hadron Distribution Amplitudes}

\author{Long-Bin Chen$^{a}$}
\email{chenlb@gzhu.edu.cn}

\author{Yao Ji$^{b,c}$}
\email{yaoji@cuhk.edu.cn}

\author{Bo-Xuan Shi$^{d}$}
\email{shibx@pku.edu.cn}

\author{Yu-Ming Wang$^{e}$}
\email{wangyuming@nankai.edu.cn}

\affiliation{
\vspace{0.2 cm}
${}^a$ School of Physics and Materials Science, Guangzhou University, Guangzhou 510006, China
\\
${}^b$ School of Science and Engineering, The Chinese University of Hong Kong, Shenzhen, Guangdong 518172, China
\\
${}^c$  Southern Center for Nuclear-Science Theory, Institute of Modern Physics, Chinese Academy of Sciences,
Huizhou, Guangdong 516000, China
\\
${}^d$  Center for High Energy Physics, Peking University, 5 Yiheyuan Road, Beijing 100871, China
\\
${}^e$  School of Physics, Nankai University, Weijin Road 94, Tianjin 300071, P.R. China
\vspace{0.2 cm}
}

\date{\today}

\begin{abstract}
\noindent

We determine for the first time the complete three-loop renormalization-group evolution kernels for flavor-nonsinglet nonforward distribution amplitudes by taking advantage of the position-space renormalization formalism and the conformal symmetry technique.
Implementing the appropriate Fourier transformations with distinct kinematic constraints then enables us to analytically construct the momentum-space renormalization kernels for both the light-flavor hadron distribution amplitudes and generalized parton distributions at three loops independently, which turn out to be fully expressible in terms of  Goncharov polylogarithms up to transcendental weight four. Applying the thus  derived   three-loop evolution equations for  the leading-twist chiral-even generalized parton distributions
further permits  us to demonstrate manifestly  the  renormalization-group invariance of the hard-collinear factorization  formulae
for the  vector flavor-nonsinglet  Compton form factors at next-to-next-to-next-to-leading-order accuracy.
\\[0.4em]

\end{abstract}


\maketitle

%
\section{Introduction}
%

Unraveling the tomographic structure of composite hadrons from the first-principles theory of strong interactions---Quantum Chromodynamics (QCD)---stands as one of the most fundamental and compelling objectives of modern hadron physics.
Elucidating the detailed quark and gluon distributions inside the high-profile nucleon and pion systems from hard scattering processes has therefore become the central focus of worldwide experimental physics programs ever since the advent of QCD.
The exclusive two-photon production of an energetic pseudoscalar meson $\pi^{0}$ at the light-like distance \cite{Lepage:1979zb,Lepage:1980fj,Braaten:1982yp,Agaev:2010aq,Wang:2017ijn} appears to be especially advantageous to explore the intriguing partonic landscape of this lightest hadron in a theoretically clean fashion and, in the meanwhile to decipher the infrared structure of perturbative non-Abelian gauge theories. The collinear QCD dynamics of such an elementary hard  exclusive reaction can be encapsulated in the universal light-cone distribution amplitude of the $\pi^{0}$-meson under the leading power approximation in the large momentum expansion, which further constitutes an indispensable ingredient for the systematic  description of the pion electromagnetic form factor \cite{Lepage:1979zb,Lepage:1980fj,Efremov:1978rn,Efremov:1979qk,Duncan:1979hi,Duncan:1979ny}
and of numerous heavy-flavor hadron decay observables \cite{Beneke:1999br,Lu:2022kos,Beneke:2001at,Ali:2007ff,Wang:2008rk,Huang:2024xii,Huang:2026mrf,Li:2010nn,Li:2012nk,Li:2012md,Lu:2009cm,
Belyaev:1994zk,Khodjamirian:2011ub,Khodjamirian:2020mlb} within the QCD-based formalisms.
It is evident that advancing the field-theoretical computations for these benchmark exclusive quantities cannot be accomplished  without controlling the renormalization-scale dependence of the leading-twist pion distribution amplitude,
which can in turn be determined  from the Efremov-Radyushkin-Brodsky-Lepage (ERBL)  evolution equation \cite{Efremov:1979qk,Lepage:1979zb}.

In parallel,  the comprehensive description of hadron structure in both longitudinal momentum and transverse spatial dimensions
has been developed rapidly  with the introduction of generalized parton distributions (GPDs) \cite{Muller:1994ses,Ji:1996ek,Ji:1996nm},
which further  open  up  an  exciting opportunity  to access  the QCD energy-momentum tensor, angular momentum,
and intrinsic  mechanical properties of hadrons \cite{Diehl:2003ny,Belitsky:2005qn,Guo:2026swz}.
These non-perturbative hadronic  distributions can be naturally probed in deeply virtual Compton scattering (DVCS)
and deeply virtual meson production (DVMP)  by virtue of hard-collinear  factorization theorems \cite{Collins:1998be,Ji:1996nm,Bauer:2002nz,Schoenleber:2024ihr,Collins:1996fb,Radyushkin:1996ru},
which express the considered scattering amplitudes in terms of perturbatively calculable short-distance coefficient functions (CFs) and universal long-distance hadronic matrix elements for a sufficiently off-shell photon state.
Interestingly, the desired renormalization-group (RG) evolution kernel of the leading-twist nucleon GPDs in the flavor-nonsinglet sector can be readily inferred from the standard ERBL kernel governing the QCD evolution of the light-flavor meson light-cone distribution amplitude (LCDA) (see, however, \cite{Beneke:2021pkl} for discussions on novel features of the ${\rm QCD \times QED}$ evolution).
This  insightful observation on two apparently disparate  evolution  kernels in  momentum  space can  ultimately  be traced back to
the identical light-cone collinear nonsinglet twist-two operator defining their position-space matrix elements.
%
It thus proves highly beneficial to construct not only the elegant and unified position-space representation of the  evolution kernel
for the twist-two nonsinglet collinear quark operator, but also its momentum-space counterparts entering the evolution  equations of the leading-twist pion LCDA and the vector  flavor-nonsinglet GPDs readily applicable to precision QCD phenomenological studies.

As a matter of fact,  the complete next-to-leading-order (NLO) QCD  computation  of the nonsinglet twist-two momentum-space ERBL  kernel had already been accomplished in both the light-cone gauge and covariant gauges \cite{Sarmadi:1982yg,Dittes:1983dy,Katz:1984gf,Mikhailov:1984ii}, with  the traditional  diagrammatic method,
more than four decades ago (see \cite{Belitsky:1999gu} for an attractive and delicate NLO reconstruction
by exploiting conformal symmetry breaking in physical (integer) dimensions \cite{Mueller:1991gd,Mueller:1993hg,Mueller:1997ak,Belitsky:1997rh}).
As mentioned above, the  necessary  NLO  nonsinglet  GPD  evolution  kernel  in momentum space  can  be algebraically
extracted from an extension procedure of the previously determined two-loop ERBL kernel without requiring additional diagrammatic calculations \cite{Dittes:1988xz,Muller:1994ses,Belitsky:1999hf},  thanks to the holomorphic properties of its restricted Fourier transform.
An alternative and innovative technical framework for deriving the position-space evolution kernel of the nonsinglet light-cone QCD operator at higher orders has been systematically developed in \cite{Braun:2014vba}, by capitalizing on the {\it exact} conformal invariance of a modified QCD theory in non-integer dimensions at the critical point where the $\beta$-function vanishes \cite{Braun:2013tva}.
Very recently, we have further  witnessed particularly encouraging progress on computing  the next-to-next-to-leading-order (NNLO) radiative corrections to a  wide range  of hard exclusive processes  with the perturbative factorization formalism,
including  the  simplest exclusive  transition $\gamma^{\ast}  \, \gamma \to \pi^{0}$ form factor ~\cite{Gao:2021iqq,Braun:2021grd},
the golden pion electromagnetic form factor \cite{Chen:2023byr,Ji:2024iak},
the  complete  DVCS    channels \cite{Braun:2020yib,Braun:2021grd,Braun:2022bpn,Ji:2023xzk},
the double deeply virtual Compton scattering (DDVCS) in the  flavor-nonsinglet vector sector \cite{Braun:2024srt},
the longitudinal  pion  DVMP process   $\gamma^{\ast}_{\|} N \to \pi N$ \cite{Chen:2026vff},
and the meson-induced exclusive Drell-Yan process \cite{Jia:2026ufq}.
As a consequence,  extending the currently available full NLO  ERBL and nonsinglet GPD evolution kernels to the  NNLO (three-loop) accuracy will be in high demand, on the one hand,  for materializing the truly complete next-to-next-to-next-to-leading-logarithmic (${\rm N^3LL}$) QCD predictions of the foregoing canonical exclusive reactions, and on the other hand, for demonstrating the anticipated factorization-scale invariance of these measurable quantities at three-loop order perturbatively.
It is our  primary  objective to fill such an important and long-overdue gap in this Letter,
by  extracting the complete and analytic position-space renormalization kernel for the twist-two flavor-nonsinglet  light-cone operator
in nonforward kinematics~\cite{Braun-3-loop-result-Comments} with the extensive use of fundamental ingredients determined from the critical conformal invariance strategy~\cite{Braun:2016qlg,Ji:2023tgu} and the three-loop nonsinglet forward anomalous dimensions~\cite{Moch:2004pa}.
In particular,  we  will continue to  provide explicitly  the  momentum-space ERBL and GPD evolution kernels at three loops
in an analytic fashion, which are evidently  indispensable for achieving high-precision phenomenological predictions of
the flagship hard  exclusive reactions accessible at the current and forthcoming  collider experiments,
including the Jefferson Lab  12 GeV upgrade~\cite{Arrington:2021alx},  the Electron-Ion Collider~\cite{AbdulKhalek:2021gbh},
and the electron-ion collider in China~\cite{Anderle:2021wcy}.
\section{General Analysis}

We  are interested in  the $\widebar{\rm MS}$-scheme evolution kernel of the flavor-nonsinglet twist-two light-cone operators
in massless QCD~\cite{Balitsky:1987bk}
\begin{eqnarray}
{\cal O}_V(z_1,z_2) &=& \bar q(z_1n) \, \gamma^{+} \, [z_1n,z_2n]q(z_2n) \,,
\notag\\
{\cal O}_A(z_1,z_2) &=&  \bar q(z_1n) \, \gamma^{+} \, \gamma^5 \, [z_1n,z_2n]q(z_2n) \,,
\label{def:tw2-O}
\end{eqnarray}
where $n^2=0$ and the straight Wilson line $[z_1n,z_2n]$ ensures gauge invariance.
A traceless flavor projection is understood for the nonsinglet kernel,
and we have defined the plus-notation $\gamma^+= n_{\mu} \, \gamma^{\mu}$.
The twist-two position-space LCDA $\Phi(z_1,z_2;\mu)$ for the pion, or more generally flavor-nonsinglet pseudoscalar mesons
composed of light quarks, is then defined as
\begin{eqnarray}
&& \langle \pi(p)|{\cal O}_A(z_1,z_2; \mu)|0\rangle  = - i \, f_\pi \, p^+  \, \Phi(z_1,z_2;\mu) \,,
\nonumber \\
&& \Phi(z_1,z_2;\mu) = \int_0^1dx\, {\rm exp} \left [ i \, p^+ \, (x z_1+\bar x z_2) \right ] \, \phi(x,\mu) \,,
\hspace{0.6 cm}
 \label{eq:pion-fourier}
\end{eqnarray}
where $\bar x = 1-x$ and we have made the scale $\mu$ dependence of the renormalized operator explicit.
In the naive dimensional regularization (NDR) scheme, which we follow throughout this work,
$\gamma^5$ in $d=4-2\epsilon$ anticommutes with all Dirac matrices.
It is then evident  that ${\cal O}_V$ and ${\cal O}_A$ share the same evolution kernel
in agreement with other non-dimensional regularization prescriptions.

By contrast, the GPDs are defined in terms of the appropriate  nonforward hadronic matrix elements of light-cone operators
with the initial and final hadrons carrying two distinct four-momenta $p$ and $p^{\prime}$.
For the nucleon target $N$, it is convenient to introduce the standard kinematic notation:
$P=(p+p^{\prime})/2$,  $\Delta=p^{\prime}-p$, the skewness parameter $\xi=-\Delta^+/(2 P^+)$  and $t=\Delta^2$.
The nonsinglet vector GPDs $H_q$ and $E_q$ are then defined as
\begin{eqnarray}
&& \hspace{-0.6 cm} P^{+} \int \frac{d z}{2 \, \pi} e^{ i x P^{+} z}
\left \langle N(p^{\prime}) \left |{\cal O}_V \left (-{z \over 2}, {z \over 2};  \mu \right) \right |N(p) \right \rangle
 \\
&& \hspace{-0.6 cm} = \overline{N}(p^{\prime})  \left[\gamma^+  H_q (x, \xi, t; \mu)
+ \frac{i\sigma^{+\nu}  \Delta_\nu}{2M}   E_q (x, \xi, t; \mu)\right]  N(p) \,,
\nonumber
\label{eq:gpd-decomposition}
\end{eqnarray}
where  $M$ denotes  the nucleon mass.
Along the same vein,  the nonsinglet axialvector GPDs can be readily  defined by the same Fourier transform
with ${\cal O}_V\mapsto {\cal O}_A$
\begin{eqnarray}
&& \hspace{-0.6 cm} P^{+} \int \frac{d z }{2 \, \pi} e^{ i x P^{+} z}
\left \langle N(p^{\prime}) \left |{\cal O}_A \left (-{z \over 2}, {z \over 2};  \mu \right) \right |N(p) \right \rangle
\\
&& \hspace{-0.6 cm} = \overline{N}(p^{\prime})  \left[\gamma^+  \widetilde{H}_q (x, \xi, t; \mu)
+ \frac{\Delta^{+}}{2M}   \widetilde{E}_q (x, \xi, t; \mu)\right] \gamma^{5} N(p)  \,.
\nonumber
\label{eq:gpd-axial-decomposition}
\end{eqnarray}

Needless to say,   the RG evolution equations of these hadronic distribution functions can be constructed from
the renormalization-scale dependence of their defining light-cone QCD  operators
\begin{eqnarray}
\frac{d \, {\cal O}(z_1,z_2;\mu)}{d \ln \mu}+ [{\mathbb H}{\cal O}](z_1,z_2;\mu) = 0 \,,
\label{eq:op_RGE}
\end{eqnarray}
where ${\mathbb H}$ is the position-space renormalization  kernel in the form of an integral operator acting on $z_1$ and $z_2$.
Such an  evolution kernel can be expanded perturbatively  in powers of the strong coupling constant
\begin{eqnarray}
{\mathbb H} = \sum_{\ell=1}^{\infty} \, a_s^{\ell} \,\,  {\mathbb H}^{(\ell)}  \,,
\qquad
a_s = {\alpha_s \over 4 \, \pi} \,.
\end{eqnarray}
Since the evolution kernels of  the  vector and axialvector  operators are identical in our renormalization scheme,
we will not distinguish between them in what follows.
The conformal properties of the light-cone  operators in  \eqref{def:tw2-O} are manifest in position space,
thereby facilitating  the bootstrap construction of the position-space representations for the renormalization kernels
of these QCD   composite operators.
On the other hand,  the short-distance matching coefficients entering  factorized scattering amplitudes are,
however,  conventionally determined in momentum space.
Consequently,  the  momentum-space evolution kernels of hadron distribution amplitudes are  more  suitable
for the  field-theoretic  description of hard exclusive reactions with the modern effective  theory paradigm.

It thus becomes  advantageous to    formulate the momentum-space evolution equations
for both the twist-two   pion  distribution amplitude   and the  vector  flavor-nonsinglet  GPDs
\begin{equation}
\boxed{
\begin{aligned}
& \frac{d  \, \phi_{\pi}(x,\mu) }{d \ln \mu}+ \int_0^1dy \, \mathbb{V}(x,y) \, \phi_{\pi}(y,\mu) = 0 \,,
 \\
& \frac{d \, F_{q}(x, \xi, t; \mu)}{d \ln \mu} + \int_{-1}^{1} dy \, \mathbb{K}(x, y, \xi) \,  F_{q}(y, \xi, t; \mu) = 0 \,,
\end{aligned}
}
\label{eq:evolution}
\end{equation}
with $F_q \in  \left \{ H_q, \, E_q \right \}$.
Subsequently,  we  can proceed to express  the position-space renormalization kernel as an integral operator \cite{Balitsky:1987bk}
\begin{eqnarray}
\boxed{[{\mathbb H} {\cal O}](z_1,z_2)=\int_0^1 d\alpha \, \int_0^1  d\beta \,
h(\alpha,\beta)  \,   {\cal O}(z_{12}^{\alpha},z_{21}^{\beta}) \,, }
\label{eq:coordinate-kernel}
\end{eqnarray}
where  $z_{12}^{\alpha}=\bar\alpha z_1+\alpha z_2$ and $\bar \alpha=1-\alpha$.
It is then straightforward to establish the desired relations between the two distinct representations
for  the evolution kernels
\begin{eqnarray}
\mathbb{V}(x,y) &=& \int_0^1 d\alpha \, \int_0^1  d\beta   \, h(\alpha,\beta) \,
\delta \left( x - \bar \alpha \,  y - \beta \, \bar y \right)  \,,
\nonumber \\
\mathbb{K}(x, y, \xi) &=& \int_0^1 d\alpha \, \int_0^1  d\beta  \, h(\alpha,\beta)   \,
\nonumber \\
&&  \, \times  \, \delta \left( x - (1-\alpha-\beta) \,  y -  (\alpha - \beta)  \, \xi \right) \,,
\label{eq:kernel-fourier-transform}
\end{eqnarray}
where  $h(\alpha,\beta)$ encodes  the support constraints, plus prescriptions and constant terms
through appropriate Heaviside step functions  and Dirac \(\delta\)-functions.

\section{Formalism}

Conformal symmetry relates the anomalous dimensions for operators involving total derivatives  to
forward anomalous dimensions and lower-loop special conformal anomalies.
This construction follows from the conformal Ward identities developed in \cite{Mueller:1991gd,Mueller:1993hg,Belitsky:1998gc,Belitsky:1998vj}.
We adopt the Wilson-Fisher (WF) critical-point formulation \cite{Braun:2013tva,Braun:2014vba,Braun:2016qlg,Braun:2017cih,Braun:2018mxm,Braun:2021tzi}
where for an arbitrary $a_s$, the parameter $\epsilon$ in $d=4-2\epsilon$ is perturbatively tuned
so that the $d$-dimensional QCD beta  function $\beta(a_s,\epsilon) = -2 \, a_s \left [ \epsilon+\bar\beta(a_s) \right ]$ vanishes.

The full conformal symmetry at the WF critical point then imposes nontrivial conditions
on the $d$-dimensional ${\mathbb H}_{\ast}$ in the form of the following commutation relations
\begin{eqnarray}
\left [S^{(\ast)}_{0, \, \pm}(a_s), {\mathbb H}_{\ast}(a_s) \right ]  = 0 \,,
\label{eq:HS-comm}
\end{eqnarray}
where $S^{(\ast)}_{0, \, \pm}(a_s)$ are $d$-dimensional conformal generators projected onto the collinear direction $n$.
They are the ${\rm SL}(2, \mathbb{R})$ group generators subject to quantum corrections
and respect the following ${\rm SL}(2, \mathbb{R})$ algebra
\begin{eqnarray}
\left [S^{(\ast)}_0(a_s), \,   S^{(\ast)}_\pm(a_s) \right ] &=& \pm \, S^{(\ast)}_\pm(a_s) \,,
\nonumber  \\
\left [S^{(\ast)}_+(a_s), \, S^{(\ast)}_-(a_s) \right ] &=& 2 \, S^{(\ast)}_0(a_s)  \,.
\end{eqnarray}
Formally, all operators in  \eqref{eq:HS-comm} are $\epsilon$-dependent.
To recover the four-dimensional evolution kernel and the associated $S_{0, \, \pm}$ in the $\widebar{\rm MS}$-scheme,
we exploit the fact that quantities in the $\widebar{\rm MS}$-scheme are $\epsilon$-independent,
and therefore, the four-dimensional operators ${\mathbb H}$ and $S_{0, \, \pm}$
can be directly obtained from their $d$-dimensional counterparts by simply trading $\epsilon$ with $-\bar\beta$
as  required by the $\beta(a_s,\epsilon) = 0$ condition.
Then, for the conformal generators, we have the following all-order expressions~\cite{Braun:2013tva,Braun:2014vba}
\begin{eqnarray}
S_0(a_s) &=& S_0^{(0)} + \left[ \bar\beta(a_s) + \frac{1}{2}{\mathbb H}(a_s) \right ] \,,
\hspace{0.2 cm}   S_- = S_-^{(0)}\,,
\nonumber \\
S_+(a_s) &=& S_+^{(0)}+(z_1+z_2)\left[ \bar\beta(a_s) +\frac{1}{2}\mathbb H(a_s) \right]
\nonumber \\
&& \, + \, z_{12} \, \Delta(a_s)\,,
\label{eq:MS-Sp}
\end{eqnarray}
where  $z_{12} = z_1 - z_2$ and the canonical generators $S_{0, \, \pm}^{(0)}$ take the following form
\begin{eqnarray}
S_+^{(0)} &=& z_1^2\partial_{z_1}+z_2^2\partial_{z_2} + 2(z_1+z_2)\,,
\notag\\
S_0^{(0)} &=&  z_1\partial_{z_1} + z_2\partial_{z_2} + 2\,,
\quad S_-^{(0)} = -\partial_{z_1}-\partial_{z_2}\,,
\hspace{0.5 cm}
\end{eqnarray}
with $\partial_{z_i} \equiv \partial/\partial z_i$.
The  conformal anomaly  $\Delta$  can be computed order  by order in perturbation theory~\cite{Braun:2016qlg}.
Since  $S_-$ corresponds to  the translation generator along the $n$-direction,  it does not receive  quantum corrections.

It turns out that  only the commutation relation  $[S_+(a_s), \,  {\mathbb H}(a_s)]=0$  imposes  nontrivial constraints
on the evolution kernel,   whose solution can be written in the form of a similarity transformation,
\begin{eqnarray}
{\mathbb H}(a_s) = \e^{-\mathbb{X}(a_s)}  \,  {{\mathbf{H}}}(a_s) \, \e^{\mathbb{X}(a_s)}  \,,
\label{eq:MSHtoCSH}
\end{eqnarray}
where ${{\mathbf{H}}}(a_s)$ is the evolution kernel in the conformal scheme (CS). 
Apparently, $\mathbb{X}$ is also an integral operator in the form of \eqref{eq:coordinate-kernel}
such  that to all orders in $a_s$
\begin{eqnarray}
[S_-^{(0)}, \, \mathbb{X}] = [S_0^{(0)},  \mathbb{X}] =0  \,.
\end{eqnarray}
In other words, both ${\mathbb H}$ and ${{\mathbf{H}}}$ commute with the canonical generators  $S_-^{(0)}$ and $S_0^{(0)}$.
Hereafter, we suppress the $a_s$ dependence for simplicity.
The operator $\mathbb{X}$ can then be determined by requiring that the $\Delta$-dependence in $S_+$ is removed by
carrying  out exactly the same similarity transformation \cite{Braun:2017cih}.
The thus-obtained  ``covariant" conformal generators
\begin{eqnarray}
{\mathbf S}_-  &=& \e^{\mathbb{X}} \, S_- \, \e^{-\mathbb{X}}  = S_-^{(0)}\,,
\nonumber  \\
{\mathbf S}_0 &=& \e^{\mathbb{X}} \, S_0 \, \e^{-\mathbb{X}}
= S_0^{(0)} + \left(\bar\beta(a_s) + \frac{1}{2}{\mathbf H}\right)  \,,
\nonumber  \\
{\mathbf S}_+ &=&  \e^{\mathbb{X}} \, S_+ \, \e^{-\mathbb{X}}
 = S_+^{(0)}+(z_1+z_2)\left(\bar\beta+\frac{1}{2}{\mathbf{H}}\right)  \,,
\hspace{0.5 cm}
\label{eq:conformal-generators}
\end{eqnarray}
and the evolution kernel $\mathbf{H}$  take on fully iterative forms requiring no additional inputs
while still satisfying the same commutation relations, 
\begin{eqnarray}
[{\mathbf S}_{0, \, \pm},   \,  {{\mathbf{H}}}] = 0 \,.
\end{eqnarray}
Comparing  \eqref{eq:conformal-generators} and  \eqref{eq:MS-Sp} order by order in  $a_s$
then  yields  a set of algebraic equations  for $\mathbb{X}$ \cite{Braun:2017cih}
\begin{eqnarray}
\left [S_+^{(0)},  \mathbb{X}^{(1)} \right] &=& z_{12} \, \Delta^{(1)}  \,,
\nonumber\\
 \left [S_+^{(0)},  \mathbb{X}^{(2)} \right ]  &=&  z_{12} \, \Delta^{(2)} + \frac{1}{2} \, [\mathbb{X}^{(1)},  \, z_{12}\Delta^{(1)}]
\nonumber  \\
&&  + \,  [\mathbb{X}^{(1)}, \,  z_1+z_2] \, \left( \beta_0 + \frac{1}{2} \, {\mathbf{H}}^{(1)} \right)  \,.
\hspace{0.5 cm}
\label{eq:anomaly-to-X}
\end{eqnarray}
Higher-order equations  follow the same recursive pattern, involving only $\Delta$
and lower-order results of ${\mathbf H}$ and $\mathbb{X}$.
The noninvariant part of $\mathbb{X}^{(\ell)}$ is fixed by \eqref{eq:anomaly-to-X},
whereas  its homogeneous part $\mathbb{X}_{\rm inv}^{(\ell)}$,
satisfying  $[S_+^{(0)},  \, \mathbb{X}_{\rm inv}^{(\ell)}]=0$, remains  arbitrary.
In the spirit of a gauge choice, we can set $\mathbb{X}_{\rm inv}^{(\ell)}=0$,
thereby fixing $\mathbb{X}^{(\ell)}$  completely.
It is therefore evident that the conformal anomaly $\Delta$  provides the information needed to
convert the evolution kernel from the CS scheme to the $\widebar{\rm MS}$ scheme
via the operator $\mathbb{X}$ appearing in  \eqref{eq:MSHtoCSH}.
The desired expressions of the integral operators $\mathbb{X}^{(1)}$ and $\mathbb{X}^{(2)}$
required to reconstruct ${\mathbb H}^{(3)}$ from ${\mathbf H}^{(3)}$ can be explicitly found in~\cite{Braun:2017cih}.

Our main task now is to determine the three-loop CS-scheme kernel ${\mathbf H}$.
This can be achieved by matching the common anomalous-dimension spectrum of ${\mathbb H}$ and ${\mathbf H}$
to the spacelike QCD  splitting function.
Since they are both integral operators of the form in  \eqref{eq:coordinate-kernel},
$z_{12}^{N-1}$ serves as an eigenfunction of both operators
\begin{eqnarray}
{\mathbf{H}}z_{12}^{N-1} = {\mathbb H}z_{12}^{N-1} = \gamma_{\widebar{\rm MS}}(N)z_{12}^{N-1} \,,
\end{eqnarray}
where $\gamma_{\widebar{\rm MS}}(N)$ stands for  the $\widebar{\rm MS}$-scheme forward anomalous dimension spectrum  \cite{Moch:2004pa}.
The second equality  can be traced back  to  the very  fact that $z_{12}^{N-1}$ is the local OPE coefficient
of twist-two operators in  \eqref{def:tw2-O} in the forward kinematics
and ${\mathbb H}$ acts as the generating function of the anomalous dimension spectrum.
In order to fully reconstruct ${\mathbf H}$ from the forward anomalous-dimension spectrum at the same order of $a_s$,
we can further  introduce the canonically invariant evolution kernel ${\mathbf H}_{\rm inv}$,
which fulfills  the defining commutation relation
\begin{eqnarray}
[S_{0, \, \pm}^{(0)},   \, {\mathbf H}_{\rm inv}] =0  \,.
\end{eqnarray}
The above equation implies that the ${\mathbf H}_{\rm inv}$ takes the following form as an integral operator,
\begin{eqnarray}
[{\mathbf H}_{\rm inv}f](z_1,z_2) = \int^1_0 d\alpha \, \int^1_0  d\beta \,\,
h_{\rm inv}(\tau) \,  f(z_{12}^\alpha,  z_{21}^\beta) \,,
\label{eq:Hinv}
\end{eqnarray}
where $f(z_1,z_2)$  denotes  an arbitrary test function and $\tau=\alpha\beta/(\bar\alpha\bar\beta)$ is the conformal ratio.
It is evident  that the eigenvalue spectrum   $\gamma_{\rm inv}(N)$  of the invariant kernel ${\mathbf H}_{\rm inv}$,
defined by  ${\mathbf H}_{\rm inv} \, z_{12}^{N-1} = \gamma_{\rm inv}(N) \, z_{12}^{N-1}$,
can be employed to fully construct the weight function $h_{\rm inv}(\tau)$ via an inverse Mellin transform.
We can then  continue to  establish  an  attractive   similarity transformation that intertwines
the CS-scheme kernel ${\mathbf H}$ with the canonically invariant kernel $\mathbf H_{\rm inv}$
as well as the conformal generators ${\mathbf S}_{0, \, \pm}$  and $S^{(0)}_{0, \, \pm}$  \cite{Ji:2023tgu}
\begin{eqnarray}
{\rm T}({\mathbf H}) \, {\mathbf H} = {\mathbf H}_{\rm inv}{\rm T} \, ({\mathbf H}) \,,
\quad
{\rm T}({\mathbf H}) \, {\mathbf S}_{0, \, \pm} =  S^{(0)}_{0, \, \pm}  \, {\rm T}({\mathbf H}) \,,
\hspace{0.5 cm}
\label{eq:Hinv-Hhat}
\end{eqnarray}
with
\begin{eqnarray}
{\rm T}({\mathbf H}) = \sum_{n=0}^\infty \, \frac{{\rm L}^n}{n!} \,
\left(\bar\beta + \frac12{\mathbf H}\right)^n  \,,
\quad
{\rm L}=\ln(z_{12}+i0) \,.
\hspace{0.5 cm}
\end{eqnarray}
Acting  both sides of the first equation in \eqref{eq:Hinv-Hhat} on  the  eigenfunction $z_{12}^{N-1}$
gives rise to  the anomalous dimension spectrum relation
\begin{eqnarray}
\gamma_{\widebar{\rm MS}} (N) = \gamma_{\rm inv}\! \left( N + \bar\beta + \frac{1}{2} \, \gamma_{\widebar{\rm MS}}(N) \right)  \,,
\label{eq:invariant-spectrum}
\end{eqnarray}
from which the canonically invariant kernel ${\mathbf H}_{\rm inv}$ has  been derived to the three-loop order~\cite{Ji:2023tgu}
using the previously determined  three-loop  forward  anomalous dimensions in Mellin-$N$ space~\cite{Moch:2004pa}.
It proves particularly convenient  to cast the first  equation of~\eqref{eq:Hinv-Hhat} in an equivalent form
\begin{eqnarray}
{\mathbf{H}} =  \mathbf H_{\rm inv} + \sum_{n=1}^{\infty} \, \frac{{\rm T}_n}{n!}
\left(\bar\beta + \frac{1}{2} \, {\mathbf{H}} \right)^n  \,,
\label{eq:all-orders-CS}
\end{eqnarray}
where the operators ${\rm T}_n$ are defined recursively
\begin{eqnarray}
\mathrm{T}_n=[\mathrm{T}_{n-1},  \, L]   \,,
\qquad   \mathrm{T}_0 = \mathbf H_{\rm inv} \,.
\label{eq:T-operators-recursion}
\end{eqnarray}
The reconstruction of the $\widebar{\rm MS}$-scheme evolution kernel at $(\ell+1)$-loop order
for  the flavor-nonsinglet twist-two operator can therefore be  schematically summarized below
\begin{empheq}[box=\fcolorbox{red}{red!20}]{align}
\gamma^{(\ell+1)}_{\widebar{\rm MS}}
\overset{\hspace*{-0.8mm}\text{\eqref{eq:invariant-spectrum}}}{\underset{{\text{Mellin}}^{-1}}{\implies}}
{\mathbf H}_{\rm inv}^{(\ell+1)}
\stackrel{\hspace*{-0.6mm}\text{\eqref{eq:all-orders-CS}}}\implies
{\mathbf H}^{(\ell+1)}
\overset{\hspace*{-0.8mm}\text{\eqref{eq:MSHtoCSH}}}{\underset{\Delta^{(\ell)}}{\implies}}
{\mathbb H}^{(\ell+1)} \,,
\label{eq:flowchart}
\end{empheq}
where  $\gamma^{(\ell+1)}_{\widebar{\rm MS}}$  denotes  the forward anomalous dimension spectrum
in the $\widebar{\rm MS}$ scheme.
Importantly, the  first two steps require the corresponding data  through $(\ell+1)$ loops,
while the last step  merely  calls for the conformal anomaly terms  up to the  $\ell$-loop accuracy.

\section{Techniques and Results}

The bootstrap procedure summarized in  \eqref{eq:flowchart} indicates that constructing the three-loop nonforward kernel
in the  $\widebar{\rm MS}$ scheme demands  a sequence of convolutions involving integral operators of the form
given in  \eqref{eq:coordinate-kernel}.
The analytical evaluation of these convolutions  turns out to be  technically rather challenging.
For illustrative purposes,  we consider  the convolution $\mathbb C_1 \otimes \mathbb C_2$
with the  two integral operators of the form  \eqref{eq:coordinate-kernel} as a basic example.
In terms of the defining representation (\ref{eq:coordinate-kernel}) for  the integral  operator of our interest
\begin{eqnarray}
[\mathbb C_i \,  f]({\bm z}) &=& \iint \limits_{{\cal D}_i} \, d\alpha_i \, d\beta_i \,
h_i(\alpha_i,\beta_i)  \,  f({{\bm M}_i {\bm z}}) \,,
\nonumber \\
{\bm z} &=&
\left(
\begin{array}{c}
z_1 \\
z_2 \\
\end{array}
\right)
\,, \qquad
{\bm M}_i=\begin{pmatrix}
\bar\alpha_i  &   \alpha_i
\\
\beta_i  &  \bar\beta_i
\end{pmatrix}
\,,
\end{eqnarray}
we can readily write down the explicit expression for the  two-fold ordered product of the integral  operators
\begin{eqnarray}
[(\mathbb C_1\otimes\mathbb C_2)f]({\bm z}) &=&  \iint \limits_{{\cal D}_1} \, d\alpha_1 \, d\beta_1 \,
\iint \limits_{{\cal D}_2} \, d\alpha_2 \, d\beta_2 \,
\nonumber \\
&& \hspace{-1.5 cm}
h_1(\alpha_1,\beta_1)  \,  h_2(\alpha_2,\beta_2)  \,   f({{\bm M}_1 \, {\bm M}_2 \,  {\bm z}}) \,.
\label{eq:coordinate-convolution}
\end{eqnarray}
This  two-fold convolution involves four integration variables, two of which must be integrated out
to  recover the standard double-integral representation for the twist-two operator  displayed  in  \eqref{eq:coordinate-kernel}.

Remarkably, these convolutions can be evaluated more conveniently in momentum-fraction space than in coordinate space.
In order to  better  understand this ``unexpected" observation, we take the construction of the momentum-space ERBL evolution kernel
as an illustrative example.
Introducing $\widehat{\mathbb C}_i$  as the counterpart of $\mathbb C_i$ in momentum-fraction space
enables us to reduce the position-space convolution shown in Eq. \eqref{eq:coordinate-convolution}
to an ordinary ordered convolution.
\begin{eqnarray}
(\widehat{\mathbb C}_1\otimes \widehat{\mathbb C}_2)(x,y)
= \int_0^1du  \,\,  \widehat{\mathbb C}_1(x,u)  \, \widehat{\mathbb C}_2(u,y) \,.
\label{eq:momentum-convolution}
\end{eqnarray}
This impressive reduction from an ordered convolution in position space to a one-dimensional convolution in momentum space
constitutes the primary reason that makes  the obtained   momentum-space kernels much simpler.
Nevertheless, we endeavor to determine the nonforward evolution kernels independently in both position and momentum space,
by capitalizing on the package \textsc{PolyLogTools}~\cite{Duhr:2019tlz} for the symbolic integration
and simplification of polylogarithmic expressions.
It is then straightforward to verify that this efficient reduction also holds for our bootstrap extraction of GPD evolution kernels,
with only a modest complication due to the additional dependence on the skewness parameter $\xi$.
We further apply the two distinct Fourier transforms associated with LCDA and GPD kinematics
to the thus-determined  three-loop position-space renormalization kernel,
thereby establishing exact agreement with the momentum-space kernels
obtained from the  direct bootstrap construction.

The eventual analytical results for both the ERBL and GPD evolution kernels in momentum space  at three-loop order
can be cast in the following schematic form (with $\mathbb{G} \in \left \{ \mathbb{V}, \, \mathbb{K} \right \}$)
\begin{eqnarray}
\mathbb{G}^{(3)} &=& C_F \, \left [ C_F^2 \, \mathbb{G}^{(3)}_{3F} + C_F C_A \, \mathbb{G}^{(3)}_{2FA}
+ C_A^2 \, \mathbb{G}^{(3)}_{F2A} \right.
\nonumber \\
&&   \left.  \,\,  +  \,  C_F \, \beta_0 \,  \mathbb{G}^{(3)}_{2F\beta_0} + \beta_1 \, \mathbb{G}^{(3)}_{F\beta_1}
+  \beta_0^2 \, \mathbb{G}^{(3)}_{F2\beta_0} \right ]\,,
\hspace{0.3 cm}
\end{eqnarray}
where the  emerged  primitive kernel functions  $\mathbb{G}^{(3)}_{3F}$,  $\mathbb{G}^{(3)}_{2FA}$
and  $\mathbb{G}^{(3)}_{F2A}$ can be expressible in terms of the genuine Goncharov polylogarithms (GPLs)
up to transcendental weight four.
We further collect explicitly  the  lengthy  expressions for the momentum-space three-loop ERBL and GPD evolution  kernels
(namely, $\mathbb{V}^{(3)}$ and $\mathbb{K}^{(3)}$)  together with the coordinate-space renormalization kernel $h^{(3)}$
in a machine-readable ancillary file attached to the arXiv submission of this  Letter.

We are now in a position to subject our newly derived three-loop flavor-nonsinglet nonforward evolution kernels
to a variety  of highly nontrivial consistency checks.

\begin{itemize}

\item{The normalization condition of the twist-two  pion  distribution amplitude  implies that the ERBL kernel
can be expressed entirely   in  the  plus-distribution form
$\mathbb{V}^{(3)}(x,y) = \left [\mathbb{P}^{(3)}(x,y) \right ]_+^x $,
which we have  verified explicitly for  our result.
The  standard  ``$+$" distribution with respect to  the variable ``$x$"   is  given by
$\int_0^1 d x  \, \left [\mathbb{P}(x,y) \right ]_+^x  \, f(x) \equiv \int_0^1 d x \, \mathbb{P}(x,y) \, \left [ f(x) -f(y) \right ]$,
where $f(x)$ is an arbitrary smooth test  function.
In addition, the coefficient of the leading singularity  of $\mathbb{P}^{(3)}(x,y)$ at $x=y$ must be proportional to
the three-loop coefficient of the universal cusp anomalous dimension  $\Gamma_{\rm cusp}$ \cite{Korchemsky:1988si}.
Under our convention, this dictates the following equation
\begin{eqnarray}
\lim_{x\to y} \, |x-y| \, \mathbb{P}^{(3)}(x,y) = -\Gamma_{\rm cusp}^{(3)} \,,
\label{eq:cusp}
\end{eqnarray}
which is in full agreement with  our result.
}

\item{Taking advantage of the  achieved  three-loop momentum-space GPD evolution kernel  $\mathbb{K}^{(3)}$
permits us to  demonstrate the desirable RG invariance of the hard-collinear factorization formulae for
the vector flavor-nonsinglet Compton form factors at three-loop accuracy
(see our companion paper~\cite{Chen:2026tbd} for further discussions).}

\item{As naturally  anticipated, taking the forward limit of the  renormalization kernel $\mathbb{K}^{(3)}$
indeed leads to the three-loop nonsinglet QCD  splitting functions  reported in \cite{Moch:2004pa}.}

\item{It has been further verified that our position-space ERBL evolution kernel $h^{(3)}$
can be reduced to the recently derived three-loop  Lange–Neubert (LN) kernel
for the leading-twist $B$-meson distribution amplitude in heavy quark effective theory (HQET)
via  the light-to-heavy reduction procedure \cite{Braun:2018fiz}
(see  \cite{Ji:2026jwu} for more comprehensive discussions).}

\item{By projecting  our  three-loop momentum-space ERBL evolution  kernel $\mathbb{V}^{(3)}$ onto the Gegenbauer basis,
we are able to fully  reproduce the  available anomalous-dimension matrix $\gamma_{m n}^{(3)}$
for $0 \leq m \leq 7$ and $0 \leq n \leq 7$ \cite{Braun:2017cih}.}

\end{itemize}

\section{Conclusions}

In summary, we have endeavored to derive for the first time the complete three-loop QCD  evolution kernels for
flavor-nonsinglet nonforward distribution amplitudes  in both momentum and position space
by taking advantage of   the conformal symmetry technique,   more than four decades after the two-loop extraction
of the renormalization kernel for the leading-twist pion LCDA~\cite{Sarmadi:1982yg,Dittes:1983dy,Katz:1984gf,Mikhailov:1984ii}.
In particular, we have further  demonstrated that the thus obtained three-loop evolution equations  of the twist-two chiral-even GPDs
can indeed  ensure  the desirable  RG invariance of the perturbative factorization formulae for the vector flavor-nonsinglet
Compton form factors at three-loop order,  thus providing an important validation
of our bootstrapped three-loop GPD  evolution kernel in momentum space.
Applying the established light-to-heavy reduction procedure to our  newly obtained  position-space evolution kernel
of the twist-two pion distribution amplitude then enabled us to recover the recently constructed  LN renormalization kernel
of the leading-twist bottom-meson distribution amplitude in HQET at three loops \cite{Ji:2026jwu}.
Interestingly,  our three-loop nonforward evolution kernels turn out to be expressible in terms of
the more general GPL function basis,  in sharp contrast with the nonsinglet QCD splitting functions
requiring only the harmonic polylogarithms   even at the four-loop accuracy \cite{Gehrmann:2026qbl}.
It can be naturally expected that the elegant and  efficient framework developed in this Letter will
open an inspiring  era of pursuing the yet higher-order QCD  predictions  for  a wide range of
the high-profile  hard  scattering processes.

%
\begin{acknowledgments}
\section*{Acknowledgements}

L.B.C. acknowledges support from the National Natural Science Foundation of China under Grant No.  12622504,
and from the Guangdong Basic and  Applied Basic Research Foundation under grant No. 2025B1515020009.
Y.J. is grateful for the support from  the National Natural Science Foundation of China with Grant No. 12535006,
from  the Guangdong Basic and Applied Basic Research Foundation under grant No. 2026A1515011306,
and from the University Development Fund of the Chinese University of Hong Kong, Shenzhen, under the grant No. UDF01003869.
B.X.S. acknowledges support from the National Natural Science Foundation of China under Grant No. 12325503.
The research of Y.M.W. is  supported in part by the  National Natural Science Foundation of China
under  Grants No. 12475097 and No. 12535006,  by the Natural Science Foundation of Tianjin under  Grant No. 25JCZDJC01190,
and by the Fundamental Research Funds for the Central Universities under   Grant No. 63261180.

\end{acknowledgments}

\section{Data Availability}

The data that support the findings of this manuscript are openly available
in the ancillary file  attached to the arXiv preprint version of this Letter.





\bibliographystyle{apsrev4-1}

\FloatBarrier
\bibliography{References}

\end{document}